\documentclass[prl,twocolumn,amsmath,amssymb,superscriptaddress]{revtex4}

\usepackage{graphicx}% Include figure files
\usepackage{dcolumn}% Align table columns on decimal point
\usepackage{bm}% bold math

\begin{document}

\title{Spinor Bose-Einstein Condensates with Many Vortices}

\author{T. Kita}
\affiliation{Division of Physics, Hokkaido University, Sapporo 060-0810,
Japan}
%\homepage{http://phys.sci.hokudai.ac.jp/~kita/index-e.html}
%\email{kita@phys.sci.hokudai.ac.jp}
\author{T. Mizushima}
\affiliation{Department of Physics, Okayama University, Okayama 700-8530,
Japan}
\author{K. Machida}
\affiliation{Department of Physics, Okayama University, Okayama 700-8530,
Japan}

\date{\today}

\begin{abstract}
Vortex-lattice structures of antiferromagnetic
spinor Bose-Einstein condensates with hyperfine spin $F\!=\! 1$ are
investigated theoretically
based on the Ginzburg-Pitaevskii equations near $T_{c}$.
The Abrikosov lattice with clear core regions are found {\em never stable}
at any rotation drive $\Omega$. Instead, each component $\Psi_{i}$
$(i\!=\!0,\pm 1)$ prefers to shift the core locations from the others to
realize
almost uniform order-parameter amplitude with complicated magnetic-moment
configurations.
This system is characterized by many competing metastable structures
so that
quite a variety of vortices may be realized with a small change in external
parameters.
\end{abstract}

%\pacs{Valid PACS appear here}

\maketitle

Realizations of the Bose-Einstein condensation (BEC)
in atomic gases have opened up a novel research field in
quantized vortices as created recently
with various techniques \cite{JILA99,ENS00,MIT01,JILA01}.
Especially interesting in these systems are vortices
of spinor BEC's in optically trapped $^{23}$Na \cite{Ketterle98} and
$^{87}$Rb \cite{Barrett01},
where new physics absent in superconductors \cite{Tinkham},
$^4$He \cite{Donnelly},
and $^3$He \cite{Volovik87,Thuneberg99,Kita01},
may be found.

Theoretical investigations on spinor BEC's were started
by Ohmi and Machida \cite{OM98} and Ho \cite{Ho98},
followed by detailed studies on vortices
with a single circulation quantum \cite{Yip99,Busch99,Volovik00,Marzlin00,
Khawaja01,Martikainen01,Kurihara01,Isoshima01,Isoshima02,Mizushima02}.
However, no calculations have been performed yet on structures
in rapid rotation where the trap potential will play a less important role.
Indeed, the clear hexagonal-lattice image of
magnetically trapped $^{23}$Na \cite{MIT01}
suggests that predictions on infinite systems are
more appropriate for BEC's with many vortices.
Such calculations have been carried out
by Ho for the single-component
BEC \cite{Ho01} and by Mueller and Ho for a two-component BEC \cite{Ho02}
near the upper critical angular velocity $\Omega_{c2}$
at $T=0$.

The purpose of the present paper is to perform detailed calculations
on vortices of $F\!=\! 1$ spinor BEC's in rapid rotation
to clarify their essential features.
To this end, we focus on the mean-field high-density phase
rather than the low-density correlated liquid phase \cite{Cooper01},
and use the phenomenological
Ginzburg-Pitaevskii (or Ginzburg-Landau)
equations near $T_{c}$ \cite{GP58,GA82}
instead of the Gross-Pitaevskii equations at $T=0$.
Since fluctuations are small in the present system,
this approach will yield quantitatively correct
results near $T_{c}$.
It should be noted that the corresponding free-energy
is formally equivalent to that derived with Ho's ``mean-field quantum Hall regime''
near $\Omega_{c2}$ at $T=0$ \cite{Ho01,Ho02}, so that the results obtained
here are also applicable to that region.

{\em Model}.--- The free-energy density of an $F\!=\! 1$ spinor BEC near
$T_{c}$
may be expanded with respect to the order parameters $\Psi_{i}$
$(i = 0, \pm 1)$ as
\begin{eqnarray}
&& \hspace{-6mm} f = -\alpha \Psi_{i}^{*}\Psi_{i}
+\Psi_{i}^{*}\,\frac{\left(-i\hbar\mbox{\boldmath $\nabla$}-
M\mbox{\boldmath $\Omega$}\!\times\!{\bf r}\right)^{ 2}}{2M}\Psi_{i}\nonumber \\
&&\hspace{0mm} 
+\frac{\beta_{\rm n}}{2}\Psi_{i}^{*}\Psi_{j}^{*}\Psi_{i}\Psi_{j}
+ \frac{\beta_{\rm 
s}}{2}\Psi_{i}^{*}\Psi_{j}^{*}(F_{\mu})_{ik}(F_{\mu})_{jl}\Psi_{k}\Psi_{l}
 \, .
\label{F}
\end{eqnarray}
Here $\alpha$, $\beta_{\rm n}$ , and $\beta_{\rm s}$ are expansion
parameters,
$F_{\mu}$ $(\mu \!=\! x,y,z)$ denotes the spin operator,
$M$ is the particle mass, and summations over repeated indices are implied.
The rotation axis $\mbox{\boldmath $\Omega$}$ is taken along ${\bf z}$.
The quantities $\beta_{\rm n}$ and $\beta_{\rm s}$ are assumed to be
constant 
near $T_{c}$ with $\beta_{\rm n}\!>\! 0$, whereas $\alpha$ changes its sign
at 
$T_{c}$ with  $\alpha\!>\! 0$ for  $T\!<\! T_{c}$.
To simplify Eq.\ (\ref{F}),
we measure the length, the energy density, the angular velocity,
and the order parameter in units of
$\hbar/\sqrt{2M\alpha}$, $\alpha^{2}/\beta_{\rm n}$, $\alpha/\hbar$,
and $\alpha/\beta_{\rm n}$,
respectively. 
The corresponding free-energy density is obtained from Eq.\ (\ref{F}) by
$\alpha\!\rightarrow\! 1$, $\beta_{\rm n}\!\rightarrow\! 1$,
$\hbar^{2}/2M\!\rightarrow\! 1$,
$M\mbox{\boldmath $\Omega$}\!\rightarrow\! \frac{1}{2}\mbox{\boldmath $\Omega$}$,
and $\beta_{\rm s}\!\rightarrow\! g_{\rm s}\!\equiv\!
\beta_{\rm s}/ \beta_{\rm n}$. It thus takes a simple form
with only two parameters $(g_{\rm s},\Omega)$.
We then introduce a couple of operators by
$a\!\equiv\! \ell(\partial_{x}\!+\! i \partial_{y})/\sqrt{2}$ and
$a^{\dagger}\!\equiv\!
\ell(-\partial_{x}\!+\! i \partial_{y})/\sqrt{2}$
with $\mbox{\boldmath $\partial$}\!\equiv \!
\mbox{\boldmath $\nabla$}-\frac{i}{2}\mbox{\boldmath $\Omega$}\!\times\!{\bf r}$
and $\ell\!\equiv\!1/\sqrt{\Omega}$
which satisfy $aa^{\dagger}\!-\!a^{\dagger}a\!=\!1$.
Equation (\ref{F}) now reads
\begin{eqnarray}
f =&& \hspace{-3mm}\Psi_{i}^{*}[ (2a^{\dagger}a\!+\! 1)\Omega-1]\Psi_{i}
+\frac{1}{2}\Psi_{i}^{*}\Psi_{j}^{*}\Psi_{i}\Psi_{j}
\nonumber \\
&&+ \frac{g_{\rm 
s}}{2}\Psi_{i}^{*}\Psi_{j}^{*}(F_{\mu})_{ik}(F_{\mu})_{jl}\Psi_{k}\Psi_{l}
\, ,
\label{F2}
\end{eqnarray}
with $\Omega_{c2}\!=\! 1$, and
the free-energy is given by
\begin{eqnarray}
{\cal F} \equiv \int  f({\bf r}) \, d{\bf r} \, .
\label{calF}
\end{eqnarray}
We can find the stable structure for each $(g_{\rm s},\Omega)$
by minimizing ${\cal F}$.
We have performed extensive calculations over the whole antiferromagnetic
region
$g_{\rm s}\!\geq\!0$, where $(\Psi_{1},\Psi_{0},\Psi_{-1})\!=\!{\rm
e}^{i\theta}
{\cal U}(0,1,0)$ and
$f\!=\!\frac{1}{2}$ at $\Omega\!=\!0$ with
$\theta$ an arbitrary phase and ${\cal U}$ the spin-space rotation
\cite{Ho98}.

A major difference from superfluid $^{3}$He \cite{Kita01} lies
in the fact that terms such as $\Psi_{i}^{*}aa\Psi_{j}$ $(i\!\neq\!j)$
are absent, i.e., 
there are no gradient couplings between different components.
Hence $\Omega_{c2}$ is the same for all components,
whereas in $^{3}$He only a single component becomes finite
at $\Omega_{c2}$ to realize the polar state \cite{Kita01}.
This degenerate feature is what characterizes the present system
to bring many competing metastable structures, as seen below.

{\em Method}.--- We minimize Eq.\ (\ref{calF})
with the Landau-level-expansion method (LLX) \cite{Kita98,Kita01}
by expanding the order parameters as
\begin{equation}
\Psi_{i}({\bf r})=
\sqrt{\cal V}\, \sum_{N=0}^{\infty}\sum_{{\bf q}} c_{N{\bf q}}^{(i)}\,
\psi_{N{\bf q}}({\bf r}) \, ,
\label{expand}
\end{equation}
with $N$ the Landau-level index,  ${\bf q}$ the magnetic Bloch vector,
and $\cal V$ the system volume.
The basis functions $\{\psi_{N{\bf q}}\}$ are eigenstates of the
magnetic translation operator
$T_{\bf R}\!\equiv\!\exp[-{\bf R}\cdot(\mbox{\boldmath $\nabla$}
+\frac{i}{2}\mbox{\boldmath $\Omega$}\times{\bf r})]$,
which can describe
periodic vortex structures of all kinds \cite{Kita98}.
Its explicit expression is given by
\begin{eqnarray}
\psi_{N{\bf q}}({\bf r})\!=&& \!\!\!\!\!\!
\sum_{n=-{\cal N}_{{\rm f}}/2+1}^{{\cal N}_{%
{\rm f}}/2}\!\!\!\!{\rm e}^{i [q_{y}(y+\ell^{2}q_{x}/2)
+na_{1x}(y+\ell^{2}q_{x}-na_{1y}/2)/\ell^{2}]}
\nonumber \\
&&\times{\rm e}^{-i xy/2\ell^{2}-(x\!- \ell^{2}q_{y}
- na_{1x})^{2}/2\ell^{2}}
\nonumber \\
&&\times \sqrt{\frac{2\pi \ell/a_{2}}{2^{N}N!\sqrt{\pi }\,{\cal V} }}H_{\!
N}\!\!\left(\! 
\frac{x\!-\! \ell^{2}q_{y}\! -\! na_{1x}}{\ell}\!\right) ,
\label{basis}
\end{eqnarray}
with ${\cal N}_{{\rm f}}^{2}$ the number of the circulation quantum
$\kappa\!\equiv\! h/M$
in the system, $H_{\! N}$ the Hermite polynomial,
and ${\bf a}_{j}$ the basic vectors in the $xy$ plane
with ${\bf a}_{2}\!\parallel\!\hat{\bf y}$
and $a_{1x}a_{2}\!=\!2\pi \ell^{2}$.
Substituting Eq.\ (\ref{expand}) into
Eq.\ (\ref{calF})
and carrying out the integration
in terms of $(s,t)\!\equiv\!
(x/a_{1x},y/a_{2}\!-\! xa_{1y}/a_{1x}a_{2})$,
the free energy is transformed into a functional of
the expansion coefficients
$\{c_{N{\bf q}}^{(i)}\}$, the apex angle $\beta \!\equiv
\!\cos ^{-1}(a_{1y}/a_{1})
$, and the ratio of the two basic vectors $\rho \!\equiv \! a_{2}/a_{1}$
as ${\cal F}\!=\! {\cal F}[\{c_{N{\bf q}}^{(i)}\},\beta,\rho ]$.
For a given $\Omega$, we directly minimize ${\cal F}/{\cal V}$ with respect
to
these quantities.

{\em Search for stable structures}.--- We here sketch our strategy
to find stable structures.
To this end,
we first summarize the basic features of the conventional Abrikosov
lattice within the framework of LLX \cite{Kita98}:
(i) any single ${\bf q}\!=\!{\bf q}_{1}$ suffices,
due to the broken translational symmetry of
the vortex lattice; (ii) the triangular (square) lattice is made up of
$N\!=\!6n$ ($4n$) Landau levels ($n\!=\!0,1,2\cdots$);
(iii) more general structures can be described by $N\!=\! 2n$ levels,
odd $N$'s never mixing up since those bases have finite amplitudes
at the cores.
This Abrikosov lattice has a single circulation quantum per unit cell.

With multi-component order-parameters,
there can be a wide variety of vortices,
which may be divided into two categories.
We call the first category as ``shift-core'' states,
where core locations are different among the three components
with an enlarged unit cell.
General structures with $n_{\kappa}$ circulation quanta per unit cell
can be described by using $n_{\kappa}$ different ${\bf q}$'s,
where the unit cell becomes $n_{\kappa}$ times as large as
that of the Abrikosov lattice.
For example, structures with two quanta per unit cell
are given by choosing
$({\bf q}_{1},{\bf q}_{2})\!=\!({\bf 0},\frac{{\bf b}_{1}+{\bf b}_{2}}{2})$,
where ${\bf b}_{1}$ and ${\bf b}_{2}$ are reciprocal lattice vectors.
This possibility was already considered by
Mueller and Ho \cite{Ho02} for a two-component system and shown
to yield stable structures.
It also describes the mixed-twist lattice to be found in
$^3$He \cite{Kita01}.
The second category may be called ``fill-core'' states
with a single circulation quantum per unit cell (i.e.,
a single ${\bf q}$ is relevant).
Here the cores of the conventional Abrikosov lattice are filled in
by some superfluid components
using odd-$N$ wavefunctions of Eq.\ (\ref{basis}).
This entry of odd-$N$ Landau levels occurs as a second-order transition
below some critical angular velocity smaller than $\Omega_{c2}/3$.
It has been shown that the $A$-phase-core vortex
in the $B$-phase of superfluid $^{3}$He belong to this category
\cite{Kita01}.

We have carried out an extensive search for stable structures
with up to $n_{\kappa}\!=\!9$ circulation quanta per unit cell, including
fill-core states.
Since we are near $T_{c}$ where normal particles are also present,
we have performed the minimization without specifying the value of the
magnetic moment $M$ for the superfluid components.
However, all the stable states found below
have $M\!=\! 0$. 
Each of the three components were expanded as Eq.\ (\ref{expand})
using $n_{\kappa}$ different ${\bf q}$'s, and the free-energy (\ref{calF})
is minimized with respect to $c_{N{\bf q}}^{(i)}$, $\beta$, and $\rho$
by using Powell's method \cite{NR}.
To pick out stable structures correctly,
we calculated Eq.\ (\ref{calF}) many times starting from different
initial values for $c_{0{\bf q}}^{(i)}$, $\beta$, and $\rho$
given randomly within $0.1\!\leq \!{\rm Re}\hspace{0.3mm} c_{0{\bf q}}^{(i)}, \,
{\rm Im}\hspace{0.3mm} c_{0{\bf q}}^{(i)}\!\leq\!0.9$,
$0.1\pi\!\leq \!\beta\!\leq\!0.5\pi$, and $0.8\!\leq \!\rho\!\leq\! 3$,
respectively.
The state with the lowest energy was thereby identified
as the stable structure.
The spin quantization axis and an overall phase were fixed conveniently
to perform efficient calculations.
Thus, any structures obtained from
the solutions below with a
spin-space rotation and a gauge transformation are also stable.

{\em Instability of Abrikosov lattice}.--- The present calculations
have revealed that the Abrikosov lattice with clear core regions
is never stable at any rotation drive $\Omega$ over the entire
antiferromagnetic region $g_{\rm s}\!\geq\!0$.
Thus, any optical experiments to detect vortices
by amplitude reductions are not suitable for
the spinor vortices.

The most stable Abrikosov lattice is given by
\begin{eqnarray}
&& \Psi_{0}({\bf r})=
\sqrt{V}\, \sum_{N}  c_{N}^{(0)}
\psi_{N{\bf q}_{1}}({\bf r}) \, ,
\label{expand-01}
\end{eqnarray}
with $c_{N{\bf q}_{1}}^{(0)}$ real,
$N\!=\! 6n$, $\beta\!=\!\frac{\pi}{3}$, and $\rho\!=\! 1$.
Here the antiferromagnetic component $\Psi_{0}$ forms
a hexagonal lattice with a single circulation quantum per unit cell.
Below some critical velocity $\Omega_{\rm f}$ smaller than $\Omega_{c2}/3$,
the core regions start to be filled in by
\begin{eqnarray}
&& \Psi_{\pm 1}({\bf r})=
i\sqrt{V}\, \sum_{N'} c_{N'}^{(1)}
\psi_{N'{\bf q}_{1}}({\bf r})\, ,
\label{expand-02}
\end{eqnarray}
with $N'$ odd.
The second transition for this odd-Landau-level entry occurs
at 
$\Omega_{\rm f}\!=\!0.1497\Omega_{c2}$ and $0.0938\Omega_{c2}$
for $g_{\rm s}\!=\!0.1$ and $0.3$, respectively.

\begin{figure}
\includegraphics[width=0.45\textwidth]{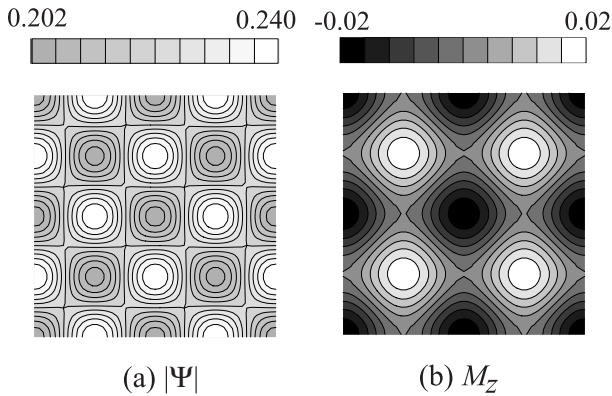}
\caption{\label{Fig1}Spatial variations of
(a) the order-parameter amplitude and
(b) the magnetic moment along $z$,
for the shift-core state (\ref{expand-11})-(\ref{expand-12})
over $|x|,|y| \! \leq\! a_{1}$
at $g_{\rm s}\!=\!0.08$ and $\Omega\!=\! 0.95\Omega_{c2}$.
The moment is directed along ${\bf z}$.}
\end{figure}

However, calculations down to $0.0001\Omega_{c2}$ of using $1800$ Landau
levels
for $g_{\rm s}\!\geq\!0$
have clarified that the above fill-core state
carries higher free energy than the following shift-core state with two
circulation quanta
per unit cell:
\begin{eqnarray}
\Psi_{1}({\bf r})=
\sqrt{V}\, \sum_{N}  c_{N}
\psi_{N{\bf q}_{1}}({\bf r}) \, ,
\label{expand-11}
\\
\Psi_{-1}({\bf r})=
\sqrt{V}\, \sum_{N} c_{N}
\psi_{N{\bf q}_{2}}({\bf r})\, ,
\label{expand-12}
\end{eqnarray}
with $c_{N}$ real and common to both,
$N\!=\! 4n$, $\beta\!=\!\frac{\pi}{2}$, $\rho\!=\! 1$, and
$({\bf q}_{1},{\bf q}_{2})\!=\!({\bf 0},
\frac{{\bf b}_{1}\!+\!{\bf b}_{2}}{2})$.
The cores of  $\Psi_{\pm 1}({\bf r})$
are shifted from each other by
$\frac{1}{2}
({\bf a}_{1}\!+\!{\bf a}_{2})$.
Figure 1 displays basic features of this shift-core state.
The magnetic moment is
ordered antiferromagnetically along $z$ axis,
and the amplitude is almost uniform taking its maximum at each site where
the moment vanishes to form the antiferromagnetic state realized in the
uniform state.

{\em Stable structures near $\Omega_{c2}$}.---
Having seen that the conventional Abrikosov lattice is never favored
in the whole antiferromagnetic domain $g_{\rm s}\!\geq\!0$,
we now enumerate all the stable structures found near $\Omega_{c2}$
to clarify their essential features.
This rapidly rotating domain is especially interesting,
because the same free energy also
becomes relevant near $\Omega_{c2}$ at $T\!=\! 0$,
as shown by Ho using a ``mean-field quantum Hall regime'' \cite{Ho01}.
Thus, the conclusions obtained here are also applicable to the
region at $T\!=\!0$.

\begin{figure}
\includegraphics[width=0.4\textwidth]{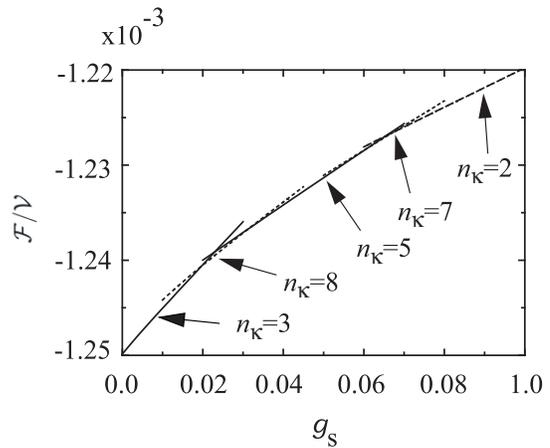}
\caption{\label{Fig2}Calculated free energy per unit volume
as a function of $g_{\rm s}$ at $\Omega\!=\!0.95\Omega_{c2}$.
Five different structures have been found for $g_{\rm s}\!\geq\!0$,
and the value of each $n_{\kappa}$
denotes the number of circulation quanta per unit cell
for each stable structure.}
\end{figure}
Figure 2 displays the lowest free energy per unit volume
as a function of $g_{\rm s}$ for $\Omega\!=\!0.95\Omega_{c2}$.
The value of each $n_{\kappa}$
denotes the number of circulation quanta per unit cell.
A special feature to be noted is that these various structures
are energetically quite close to each other;
for example, the $n_{\kappa}\!=\! 8$ state at $g_{\rm s}\!=\!0.02$
is favored over the $n_{\kappa}\!=\!3$ state
by a relative free-energy difference of order $10^{-6}$.
This fact suggests that we may realize
quite a variety of metastable structures by a small change
of the boundary conditions, the rotation process, etc.

Details of these structures are summarized as follows.

The $n_{\kappa}\!=\!2$ state of Eqs.\ (\ref{expand-11}) and
(\ref{expand-12}) 
is stable for $g_{\rm s}\geq 0.0671$. We have already seen
its basic features above.

The $n_{\kappa}\!=\!3,5,7$ states can be expressed compactly as
\begin{eqnarray}
&&\hspace{-8mm}\Psi_{0}=\sum_{N}\sum_{\nu=1}^{\frac{n_{\kappa}-1}{2}}c_{N{\bf q}_{\nu}}^{(0)}
\bigl(\,\psi_{N{\bf q}_{\nu}}
-{\rm e}^{-2i\frac{\nu}{n_{\kappa}}\pi}\psi_{N{\bf
q}_{n_{\kappa}-\nu}}\,\bigr)\, ,
\\
&&\hspace{-8mm}\Psi_{\pm
1}=\sum_{N}\Biggl[\,\sum_{\nu=1}^{\frac{n_{\kappa}-1}{2}}c_{N{\bf q}_{\nu}}^{(\pm 1)}
\bigl(\,\psi_{N{\bf q}_{\nu}}
+{\rm e}^{-2i\frac{\nu}{n_{\kappa}}\pi}\psi_{N{\bf
q}_{n_{\kappa}-\nu}}\,\bigr)
\nonumber \\
&&\hspace{11mm}+c_{N{\bf q}_{n_{\kappa}}}^{(\pm 1)}\psi_{N{\bf
q}_{n_{\kappa}}}\Biggr]\, ,
\end{eqnarray}
where $N$'s are even and ${\bf q}_{\nu} \!=\!
\frac{\nu}{n_{\kappa}}{\bf b}_{1}$.
These $n_{\kappa}\!=\!3,5,7$ states are stable for $0\!\leq\! g_{\rm
s}\!\leq\! 0.0196$,
$0.0313\!\leq\! g_{\rm s}\!\leq\! 0.0613$,
and $0.0613\!\leq\! g_{\rm s}\!\leq\! 0.0671$, respectively.
Unlike the two component system considered by Mueller and Ho \cite{Ho02}
where each component is specified by a single-${\bf q}$ basis function,
$\Psi_{m}$ here is made up of multiple
basis functions $\psi_{N{\bf q}_{\nu}}$ whose cores are shifted from each
other by 
$(\nu/n_{\kappa}){\bf a}_{2}$.
Figure 3 displays the basic features of the $n_{\kappa}\!=\!3$ state.
The lattice is hexagonal at $g_{\rm s}\!=\!0$, but deforms rapidly
as $g_{\rm s}$ increases. The order-parameter amplitude is again
almost constant, and the magnetic moment ${\bf M}$ has a
complicated structure.
These features are common to all the $n_{\kappa}\!\geq\!3$
states discussed here, although no details are presented for the other
states.
The lattice parameters $(\beta,\rho)$ for $n_{\kappa}\!=5,7$ states
are $(0.1311\pi,2.056)$ and $(0.1830\pi,2.609)$ for $g_{\rm s}\!=\!0.05$
and $0.066$, respectively, which change little in each relevant range
of stability.

\begin{figure}
\includegraphics[width=0.44\textwidth]{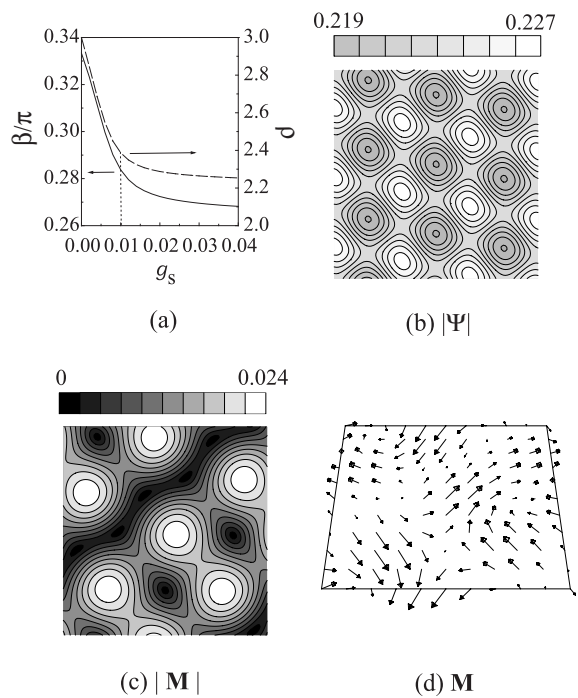}
\caption{\label{Fig3}(a) Variations of $\beta/\pi$ and $\rho$
as a function of $g_{\rm s}$ for the $n_{\kappa}\!=\!3$ state
at $\Omega\!=\! 0.95\Omega_{c2}$.
This lattice for $g_{\rm s}\!=\!0$ is hexagonal with $\beta\!=\!\pi/3$ and
$\rho\!=\!3$.
Figures 2(b)-(d) display, for $g_{\rm s}\!=\!0.01$,
spatial variations of
(b) the order-parameter amplitude,
(c) amplitude of the magnetic moment,
and (d) the magnetic moment,
over $-3a_{1x}/2<x<3a_{1x}/2$ and $-a_{2}/2<y<a_{2}/2$.
}
\end{figure}

The remaining $n_{\kappa}\!=\!8$ state, stable over $0.0196\!\leq\! g_{\rm s}\!\leq\! 0.0313$,
 is given by
\begin{eqnarray}
&&\hspace{-8mm}\Psi_{0}=\sum_{N}c_{N{\bf q}_{1}}^{(0)}
\big(\psi_{N{\bf q}_{1}}+\psi_{N{\bf q}_{2}}
-\psi_{N{\bf q}_{3}}+i\psi_{N{\bf q}_{4}}\bigr)\, ,
\\
&&\hspace{-8mm}\Psi_{\pm 1}= 
\sum_{N}\sum_{\nu=5}^{8}c_{N{\bf q}_{\nu}}^{(\pm 1)}\psi_{N{\bf q}_{\nu}} \, ,
\end{eqnarray}
where ${\bf q}_{1}= \frac{{\bf b}_{1}+{\bf b}_{2}}{2}$, ${\bf q}_{2} ={\bf
0}$,
${\bf q}_{3} =  \frac{{\bf b}_{1}-{\bf b}_{2}}{4}$,
${\bf q}_{4} =  \frac{-{\bf b}_{1}+{\bf b}_{2}}{4}$, ${\bf q}_{5} =
\frac{{\bf b}_{1}}{2}$,
${\bf q}_{6} = \frac{{\bf b}_{2}}{2}$,
${\bf q}_{7} =  \frac{{\bf b}_{1}+{\bf b}_{2}}{4}$, and
${\bf q}_{8} =  \frac{-{\bf b}_{1}-{\bf b}_{2}}{4}$.
The parameters $(\beta,\rho)$ at $g_{\rm s}\!=\!0.025$ are
$(0.3317\pi,1.049)$, and changes
only slightly in the above range of $g_{\rm s}$.

{\em Concluding remarks}.---We have performed extensive calculations
on antiferromagnetic $F\!=\! 1$ spinor vortices in rapid rotation.
The conventional 
Abrikosov lattice is shown never favored.
Each stable structure has almost constant order-parameter amplitude
and a complicated magnetic-moment configuration,
as shown in Figs.\ 1 and 3, for example.
This means that any optical experiments to detect vortices
by amplitude reduction will not be suitable for the spinor vortices.
Instead, techniques to directly capture spatial magnetic-moment configurations
will be required.
The system has many metastable structures
which are different in the number of circulation
quanta per unit cell $n_{\kappa}$,
but are quite close to each other energetically.
Thus, domains to separate different structures
may be produced easily.
This degenerate feature is also present
within the $(\beta,\rho)$ space of a fixed $n_{\kappa}$,
where $\beta$ is the vortex-lattice apex angle and $\rho$ is the length
ratio of the two basic vectors.
Put it differently, we can deform a stable lattice structure
with a tiny cost of energy.
These facts indicate that the spinor BEC's can be a rich source of novel vortices
realized with a small change in external parameters.

This research is supported by a Grant-in-Aid for Scientific Research from the
Ministry of Education, Culture, Sports, Science, and Technology of Japan.

%%%%%%%%%%%%%%%%%%%%%%%%%%%%%%%%%%%%%%%%%%%%%%%%%%%%%%%%%%%%%%%%%%%%%

%%%   The Bibliography

%%%%%%%%%%%%%%%%%%%%%%%%%%%%%%%%%%%%%%%%%%%%%%%%%%%%%%%%%%%%%%%%%%%%%


\begin{references}

\bibitem{JILA99} M. R. Matthews, B. P. Anderson, P. C. Haljan, D. S. Hall,
C. E. Wieman, and E. A. Cornell, Phys. Rev. Lett. {\bf 83}, 2498 (1999).

\bibitem{ENS00} K. W. Madison, F. Chevy, W. Wohlleben, and J. Dalibard,
Phys. Rev. Lett. {\bf 84}, 806 (2000).

\bibitem{MIT01} J. R. Abo-Shaeer, C. Raman, J. M. Vogels, and W. Ketterle,
Science {\bf 292}, 476 (2001).

\bibitem{JILA01} P. C. Haljan, I. Coddington, P. Engels, and E. A. Cornell,
Phys. Rev. Lett. {\bf 87}, 210403 (2001).

\bibitem{Ketterle98}J. Stenger, S. Inouye, D. M. Stamper-Kurn, H.-J.
Miesner,
A. P. Chikkatur, and W. Ketterle, Nature {\bf 369}, 345 (1998).

\bibitem{Barrett01}M. D. Barrett, J. A. Sauer, and M. S. Chapman,
Phys. Rev. Lett. {\bf 87}, 010404 (2001).

\bibitem{Tinkham} See, e.g., M. Tinkham, {\em Introduction to
Superconductivity} 
(McGraw-Hill, New York, 1996).

\bibitem{Donnelly} R. J. Donnelly, {\em Quantized Vortices in Helium II}
(Cambridge University Press, Cambridge, 1991).

\bibitem{Volovik87} M. M. Salomaa and G. E. Volovik, Rev. Mod. Phys. {\bf
59}, 533 (1987).

\bibitem{Thuneberg99} O. V. Lounasmaa and E. Thuneberg,
Proc. Nath. Acad. Sci. USA {\bf 96}, 7760 (1999).

\bibitem{Kita01} T. Kita, Phys. Rev. Lett. {\bf 86}, 834 (2001).

\bibitem{OM98} T. Ohmi and K. Machida, J. Phys. Soc. Jpn. {\bf 67}, 1822
(1998).

\bibitem{Ho98} T.-L. Ho, Phys. Rev. Lett. {\bf 81}, 742 (1998).

\bibitem{Yip99} S. K. Yip, Phys. Rev. Lett. {\bf 83}, 4677 (1999).

\bibitem{Busch99} Th. Busch and J. R. Anglin,
Phys. Rev. A{\bf 60}, R2669 (1999).

\bibitem{Volovik00} U. Leonhardt and G. E. Volovik, JETP Lett. {\bf 72}, 46
(2000).

\bibitem{Marzlin00} K.-P. Marzlin, W. Zhang, and B. C. Sanders,
Phys. Rev. A{\bf 62}, 013602 (2000).

\bibitem{Khawaja01} U. A. Khawaja and H. T. C. Stoof, Nature {\bf 411}, 918
(2001);
Phys. Rev. A{\bf 64}, 043612 (2001).

\bibitem{Martikainen01} J.-P. Martikainen, A. Collin, and K.-A. Suominen, cond-mat/0106301.

\bibitem{Kurihara01} S. Tuchiya and S. Kurihara, J. Phys. Soc. Jpn. {\bf
70}, 1182 (2001).

\bibitem{Isoshima01} T. Isoshima, K. Machida, and T. Ohmi, J. Phys. Soc.
Jpn. {\bf 70}, 1604 (2001).

\bibitem{Isoshima02} T. Isoshima and K. Machida, cond-mat/0201507.

\bibitem{Mizushima02} T. Mizushima, K. Machida, and T. Kita, cond-mat/0203242.

\bibitem{Ho01} T.-L. Ho, Phys. Rev. Lett. {\bf 87}, 060403 (2001).

\bibitem{Ho02} E. J. Mueller and T.-L. Ho, cond-mat/0201051.

\bibitem{Cooper01}N. R. Cooper, N. K. Wilkin, and J. M. F. Gunn, Phys. Rev. Lett. 
{\bf 87}, 120405 (2001).

\bibitem{GP58} V. L. Ginzburg and L. P. Pitaevskii, Zh. Eksp. Teor. Fiz.
{\bf 34}, 1240 (1958) [Sov. Phys. JETP {\bf 7}, 858 (1958)].

\bibitem{GA82} For a review on this approach, see,
V. L. Ginzburg and A. A. Sobyanin, J. Low Temp. Phys.
{\bf 49}, 507 (1982).

\bibitem{Kita98}  T. Kita, J. Phys. Soc. Jpn. {\bf 67}, 2067 (1998).

\bibitem{NR}  W. H. Press, S. A. Teukolsky, W. T. Vetterling,
and B. P. Flannery, {\em Numerical Recipes in C}
(Cambridge University Press, Cambridge, 1988).


\end{references}
\end{document}